\documentclass[aps,pra,tightenlines,twocolumn]{revtex4-2}
\usepackage[T1]{fontenc}
\usepackage[utf8]{inputenc}
\usepackage{amsmath}
\usepackage{amsfonts}
\usepackage{amssymb}
\usepackage{graphicx}
\usepackage[english]{babel}
\usepackage{hyperref}

\usepackage{physics}
\usepackage{bm}
\usepackage{slashed}
\usepackage{xcolor}
\usepackage{booktabs}

\newcommand{\R}{\mathbb{R}}

\DeclareMathOperator*{\argmin}{arg\,min}

\DeclareMathOperator{\dist}{dist}
\newcommand{\T}{\mathsf{T}}
\newcommand{\mcal}{\mathcal M}

\begin{document}
\title{Buckling, Crumpling, and Tumbling of Semiflexible Sheets in Simple Shear Flow}

\author{Kevin S. Silmore}
\author{Michael S. Strano}
\author{James W. Swan}
\email{jswan@mit.edu}
\affiliation{ Department of Chemical Engineering, Massachusetts Institute of Technology, Cambridge, MA, 02139, USA}

\date{April 25, 2021}

\begin{abstract}
As 2D materials such as graphene, transition metal dichalcogenides, and 2D polymers become more prevalent, solution processing and colloidal-state properties are being exploited to create advanced and functional materials. However, our understanding of the fundamental behavior of 2D sheets and membranes in fluid flow is still lacking. In this work, we perform numerical simulations of athermal semiflexible sheets with hydrodynamic interactions in shear flow. For sheets initially oriented in the flow-gradient plane, we find buckling instabilities of different mode numbers that vary with bending stiffness and can be understood with a quasi-static model of elasticity. For different initial orientations, chaotic tumbling trajectories are observed. Notably, we find that sheets fold or crumple before tumbling but do not stretch again upon applying greater shear.
\end{abstract}

\maketitle

\section{Introduction}
What do small 2D sheets do when subjected to fluid flow? Researchers have long studied problems of fluid-structure interactions ranging from flapping flags at macroscopic scales to tumbling polymer chains, such as DNA, at microscopic scales. This work follows a similar vein of research and, to our knowledge, represents one of the first studies of the dynamical behavior of athermal 2D sheets in shear flow at low Reynolds number. The quantification of different dynamical regimes explored here should help inform solution processing methods for nanomaterials and techniques to manipulate flexible materials via fluid flow. Such 2D materials of interest include graphene and graphene derivatives, clays, inorganic nanosheets, nacre-like materials, elastic membranes, colloidal membranes, 2D polymers, and 2D biological objects like kinetoplasts among others.

The dynamical behavior of semiflexible filaments and polymer chains dispersed in fluids is associated with a rich history of academic and industrial exploration. de Gennes famously studied the coil-stretch transition of polymers in extensional flow \cite{degennes1974}, and subsequent theoretical \cite{winkler2006} and experimental work \cite{schroeder2005,smith1999,harasim2013} has looked at the behavior of polymers in shear flow. In particular, it has been shown that polymers stretch under shear but tumble somewhat erratically due to thermal fluctuations. Many additional simulation works \cite{nikoubashman2016, nikoubashman2017,huang2011,dalal2012,munk2006,doyle1997} have also enhanced our understanding of the rheological properties of polymer chains. At perhaps a slightly larger length scale lies a significant body of work on fluid-structure interactions of flexible fibers and filaments \cite{lindner2015}. Namely, fibers have been found to exhibit buckling \cite{young2007,kantsler2012, quennouz2015} and morphological transitions \cite{liu2018a} as well as periodic and chaotic dynamical trajectories \cite{slowicka2015, slowicka2020, lagrone2019}.

At high Reynolds number, one example of fluid-structure interactions with \textit{2D objects} that has attracted researchers' attention is the flapping of flags \cite{shelley2011, argentina2005}. Relatively less work, though, has focused on the behavior of 2D semiflexible sheets in low-Reynolds-number flows, despite the increasing relevance of transition metal dichalogenides, graphene \cite{liu2018b}, graphene oxide \cite{tardani2018a, parviz2015}, and 2D polymers \cite{colson2013,payamyar2016} in advanced technologies. Xu and Green \cite{xu2014, xu2015} looked at sheets under shear and biaxial extensional flow, and Babu and Stark \cite{babu2011} have studied tethered sheets in fluids, confirming predicted scaling laws of Frey and Nelson \cite{frey1991} regarding thermal fluctuations in the presence of hydrodynamic interactions. Additionally, Dutta and Graham \cite{dutta2017} have studied Miura-patterned sheets and observed various interesting dynamical regimes involving periodic tumbling, unfolding, and quasiperiodic limit cycles. Most recently, Yu and Graham \cite{yu2021} studied ``compact-stretch'' transitions of elastic sheets under extensional flow. However, there is still much to be learned about the dynamics of sheets subjected to fluid flow. Such fundamental understanding is particularly relevant now, as solution processing and liquid exfoliation \cite{stafford2018, ambrosi2018, coleman2011, nicolosi2013} are commonly employed techniques in handling 2D materials, and nanosheet mechanical properties can be tuned by altering exfoliation protocols (e.g., solvent properties, the use of surfactant, etc.) \cite{nicolosi2013,varrla2015}. Furthermore, studies involving 2D materials whose functions depend significantly on coupled fluid motion are becoming more prevalent \cite{mallory2015, gibaud2017a, ding2017, tardani2018a, davidson2018a, laskar2018, kamal2020, pezzulla2020, klotz2020}.

\begin{table}[h]
\centering
\small \caption{Approximate values of bending rigidities for some common materials and nanomaterials}
\label{tab:energies}

\begin{tabular}{@{}ll@{}}
\toprule 
Material & Bending Rigidity ($\left. k_B T \right|_{\text{298 K}}$) \\ 
\midrule 
Graphene & $40$ \cite{nicklow1972}, $280$ \cite{lindahl2012}, $10^5$ \cite{blees2015}  \\
Graphene oxide & $1$ \cite{poulin2016} \\
MoS$_2$ & $370$ \cite{jiang2013} \\
Phospholipid bilayer & $20$ \cite{seifert1997} \\
PMMA (100 nm film) & $8 \times 10^7$ \cite{brandrup1999} \\
\bottomrule 
\end{tabular} 

\end{table}

We consider asymptotically thin, isotropic, semiflexible sheets of size $L$ immersed in a fluid at low Reynolds number. Sheets can be characterized by a bending rigidity, $\kappa$, and a 2D Young's modulus, $Y$, that have units of energy and force per length, respectively. The dimensionless ratio relating these two quantities is known as the Föppl-von Kármán (FvK) number and is equal to $YL^2/\kappa$. From the classical theory of thin plates \cite{landau2008}, both $\kappa$ and $Y$ are also related to the 3D Young's modulus of the material, $E$, as $Y = Eh$ and $\kappa =  Yh^2 / [12(1-\nu^2)]$, where $h$ is the thickness of the sheet and $\nu$ is the Poisson ratio.
As reference, some examples of bending rigidities are provided in Table \ref{tab:energies}.
For graphene, values can vary significantly from the theoretical microscopic bending rigidity (the first value) due to thermal fluctuation-induced stiffening, which is, in turn, experimentally challenging to measure and length-scale dependent \cite{nelson1987, bowick2017}. It is also worth noting that many nanomaterials are produced containing multiple layers, and it has been shown that multilayer van der Waals materials exhibit intermediate behavior between classical thin plates and ideally lubricated stacks of sheets \cite{wang2019}. While these complications and others, such as nonzero slip \cite{kamal2020}, are present for atomically thin nanomaterials, an athermal semiflexible membrane model may at least qualitatively describe the behavior of sheets with lateral dimensions much larger than the atomic scale and with (renormalized) bending rigidities much greater than the thermal energy (i.e., $\kappa / k_B T \gg 1$).

In this work, we quantify the behavior of high-FvK sheets subjected to shear flow using a numerical immersed boundary method. For different ratios of bending rigidity to shear energy (a dimensionless ratio we denote as $S$) and initial orientations, we find behavior ranging from buckling to transient tumbling and chaotic crumpling. Specifically, for initially flat sheets oriented near the flow-vorticity plane, quasi-1D buckling reminiscent of Euler buckling is observed, and a simple continuum elasticity model is presented to explain transitions in the buckling modes. The orientation of sheets that are stiff (relative to the shear strength) or initially oriented near the flow-vorticity plane is found to be well predicted by Jeffery's equations for thin oblate spheroids. However, deviations from the Jeffery orbits are observed for different initial orientations, with sheets of intermediate bending rigidity transiently tumbling before following a Jeffery-like trajectory and sheets of low bending rigidity crumpling into a compact structure and continuously tumbling in a chaotic manner.
Summary statistics of sheet orientation (viz., the mean orientation and the orientational covariance matrix) are constructed and used to analyze the crumpling and chaotic motions quantitatively.

\section{Model and Methods}

Hexagonal sheets of circumradius $58a$ were constructed by creating a surface triangulation with edges of length $l=2a$ and placing beads of radius $a$ at each of the vertices (for a total of $N=2611$ beads).
These beads can be considered a geometric manifestation of the shortest resolvable length scale, $a$, in this coarse-grained sheet model. We caution readers to avoid thinking of $a$ as a thickness since the sheets in this study behave hydrodynamically as though they are asymptotically thin, as discussed more below.
We chose to use hexagonal sheets due to their symmetry and the density of the underlying triangular lattice of beads. Such a packing of beads is the closest possible in 2D, which is important for capturing self-avoidance and uniform hydrodynamic interactions, and represents a high degree of rotational symmetry, which makes the barrier to bending as a function of angle relatively uniform. These choices reflect a desire to create a discrete sheet model that tries to be as ``isotropic as possible''.

Bending forces were modeled via dihedral forces over each pair of neighboring triangles $\bigtriangleup_i$ and $\bigtriangleup_j$ as:
\begin{equation}
U_\mathrm{bend}(\bigtriangleup_i, \bigtriangleup_j) = \kappa (1 - \vu n_i \vdot \vu n_j),
\end{equation}
where $\kappa$ is the bending rigidity and $\vu n_i$ and $\vu n_j$ are (consistently oriented) triangle normals \cite{bowick2017, bian2020, guckenberger2017, gompper1996}. Note that such a model of bending may not be generally applicable for capturing bending forces of real materials (e.g., those that are nonlinear or plastically deform) or other models (e.g., the Helfrich-Canham model) at large curvatures. However, it is particularly desirable for efficiency and for capturing deviations from flat conformations. The equivalent continuum bending rigidity is given by $\tilde \kappa = \kappa / \sqrt{3}$ \cite{gompper1996}.

Hard-sphere interactions between all of the beads (excluding those connected by a bond) were approximated via the pair potential,
\begin{equation}
U_\mathrm{HS}(r) = \begin{cases}
\frac{16 \pi \eta  a^2 \qty[2a\ln(\frac{2a}{r}) + (r - 2a)]}{\Delta t} & 0 \leq r \leq 2a \\
0 & r > 2a
\end{cases},
\end{equation}
where $r$ is the distance between two interacting particles, and $\Delta t$ is the integration timestep used. This pairwise potential displaces two overlapping particles to contact under Rotne-Prager-Yamakawa dynamics (discussed below) and accurately captures the thermodynamic properties of the hard sphere fluid \cite{heyes1993}.

Harmonic bonds of the form
\begin{equation}
U_\mathrm{bond}(r) = \frac{k}{2} (r - r_0)^2
\end{equation}
were applied along each of the edges of the triangulation with stiffness $k = 1000 \times 6\pi\eta\dot \gamma a$ and $r_0 = 2a$, where $\eta$ and $\dot \gamma$ are the viscosity and shear rate of the surrounding fluid. The continuum 2D Young's modulus is related to $k$ via the expression $Y = 2 k / \sqrt{3}$ \cite{bowick2017}. With these values of bond stiffness and bending rigidity, the Föppl-von Kármán (FvK) number of the system ranged in magnitude between $10^4$ and $10^6$, the latter of which is similar to that of a sticky note or a 100 nm-wide sheet of graphene. Thus, with these high FvK numbers and by scaling the in-plane stiffness with the flow strength, the sheets studied here can largely be considered \textit{inextensible}, and we expect in-plane modes of motion to be generally irrelevant to the behavior observed.

\begin{figure*}[ht]
\centering
\includegraphics[width=6in]{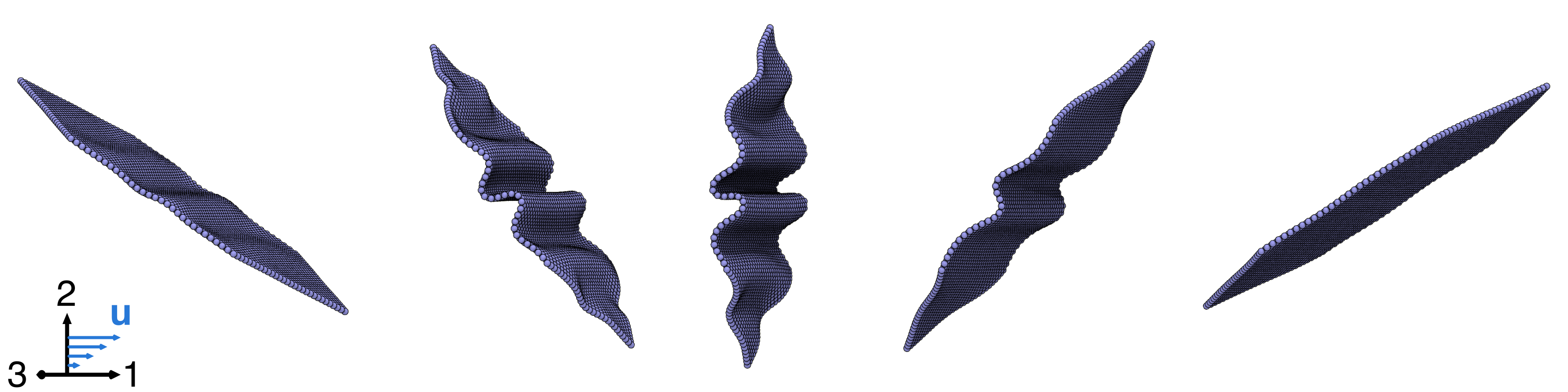}
\caption{Snapshots of a buckling sheet with initial angle $\phi = 0^\circ$ and dimensionless bending rigidity $S = 4.61 \times 10^{-4}$ at times $\dot \gamma t = 10$,  $10.75$, $11.5$, $12.25$, and $13$ from left to right. The axis labels 1, 2, and 3 represent the flow, gradient, and vorticity directions, respectively.}
\label{fig:buckle_images}
\end{figure*}

For the surrounding fluid, we assume a low Reynolds number such that Stokes equations are valid. Hydrodynamic interactions between all of the beads in the sheet were included at the Rotne-Prager-Yamakawa (RPY) level \cite{rotne1969, yamakawa1970}. That is, each bead produces a Stokes monopole and degenerate quadrupole but is ``regularized'' in such a way that ensures the mobility tensor remains positive definite for all (potentially overlapping) particle configurations. Although ostensibly a low-level of approximation for hydrodynamic interactions, it is important to note that the beads of the sheet act as regularized Stokeslets, so, in some sense, the use of large sheets with many beads employed here can be considered an approximation of the appropriate boundary integral for a smooth, continuous sheet \cite{fiore2017,swan2016}. Each $3 \times 3$ block of the RPY tensor mapping forces on particle $j$ to the velocity of particle $i$ is a function of the distance, $r$, between the particles and is given by:
\begin{equation}
\bm{\mcal}_{ij} = 
\frac{1}{6 \pi \eta a}
\begin{cases}
\qty(\frac{3a}{4r} + \frac{a^3}{2r^3}) \vb I + \qty(\frac{3a}{4r} - \frac{3a^3}{2r^3})\vu r \vu r^\T & r > 2a \\
\qty(1 - \frac{9r}{32a}) \vb I + \frac{3r}{32a} \vu r \vu r^\T & r \leq 2a
\end{cases},
\end{equation}
where $\vu r$ is a unit vector pointing from particle $i$ to particle $j$.
The particles were then advanced in time by integrating the following equation of motion:
\begin{equation}
\dd{ \vb x_i} = \qty(-\sum_j \bm{\mcal}_{i j} \pdv{U}{\vb x_j} + \vb L  \vb x_i  )\dd{t},
\label{eq:main_int}
\end{equation}
where $\vb x \in \R^{3N}$ is a stacked vector of $N$ particle coordinates, $U$ is the total potential energy and $(\vb L)_{mn} = \dot \gamma\delta_{m1}\delta_{n2}$ is the $3 \times 3$ velocity gradient tensor. Importantly, advecting the particles as ``no-slip'' point particles and not as rigid spheres (which would require more sophisticated integrators that operate at least at the stresslet level \cite{fiore2018} and include translational-dipolar couplings) causes the sheet to behave as if it were asymptotically thin with respect to the shear flow in the sense that $h/L \to 0$, where $h$ is the thickness. This lack of thickness breaks the periodicity of Jeffery orbits and introduces stable flat states in the flow-vorticity plane, but otherwise does not significantly affect the trajectories away from these stable states compared to a very thin sheet. In fact, it is easy to show that the \textit{inverse} of the Jeffery orbit period, $T$,  of an oblate spheroid scales with the aspect ratio as $T^{-1} \approx \frac{\dot \gamma}{2 \pi}\frac{h}{L}$ for small values of $h/L$, so the periods of thin sheets can become arbitrarily long. Thus, we believe the phenomena reported here should be generalizable to more realistic sheets of small but nonzero aspect ratios.

Incorporating higher-order hydrodynamic interactions \cite{fiore2018, wajnryb2013, kim2005} poses several theoretical and computational challenges, the most important of which is the necessity to develop an elasticity model that properly constrains the relative rotation of beads within a sheet. Development of such a model in future work could be valuable.

\begin{figure}[h]
\centering
\includegraphics{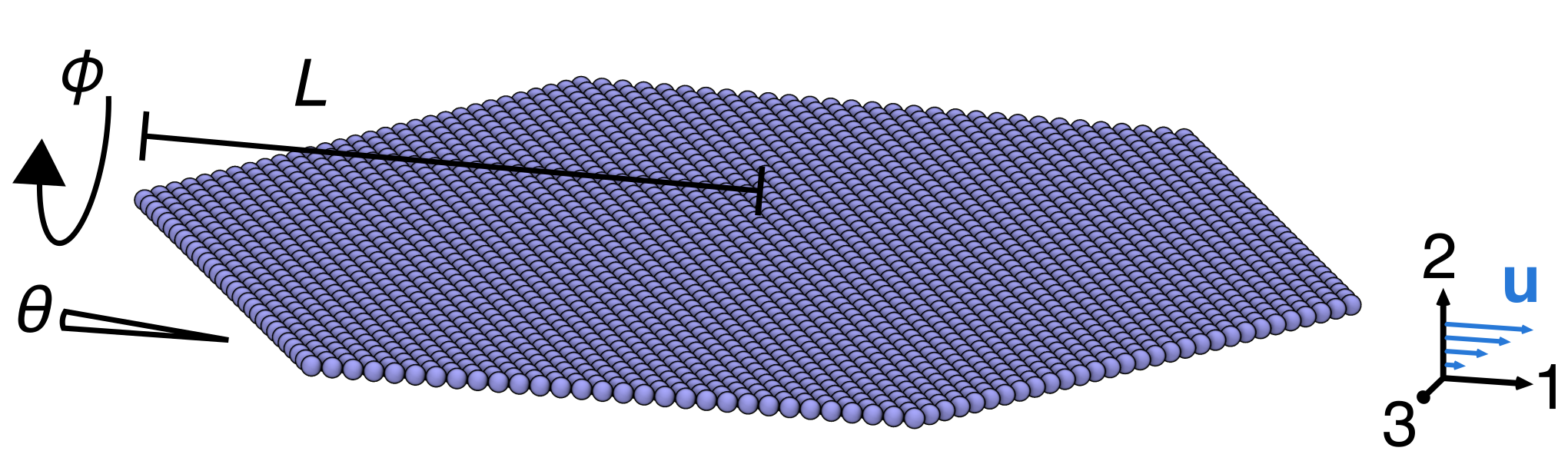}
\caption{Illustration of the initial angles, $\theta$ and $\phi$, used for the initial conditions of the sheet as well as the length of the sheet, $L$, and the flow (1), gradient (2), and vorticity (3) directions of the imposed shear flow.}
\label{fig:angle_diagram}
\end{figure}

Flat sheets initially constructed in the flow-vorticity plane were rotated by a varying angle $\phi$ about the flow axis and then by an angle $\theta = 5^\circ$ about the vorticity axis (see Figure \ref{fig:angle_diagram}). Rotating by $\theta$ perturbs the sheet away from the stable state $(\theta, \phi) = (0^\circ, 0^\circ)$. By symmetry, unique initial conditions are only generated by angles $\phi \in [0^\circ, 90^\circ]$. Simulations were conducted by integrating equation \ref{eq:main_int} via a forward Euler scheme with a timestep of at most $\dot \gamma \Delta t = 5\times10^{-4}$ using a custom plugin adapted from ref. \cite{fiore2017} for the HOOMD-blue molecular simulation package \cite{anderson2020} on graphics processing units (GPUs). This timestep was chosen to ensure numerical stability and was sufficiently small compared to the highest-frequency mode of motion in the system (i.e., the harmonic bonds between the beads). All simulations were conducted on NVIDIA GTX 1080 Tis and required thousands of GPU-hours.

\section{Buckling}

The dynamics of rigid ellipsoidal particles in simple shear flows was first worked out by Jeffery \cite{jeffery1922, kim2005} with details for non-axisymmetric particles later studied by Hinch and Leal \cite{hinch1979}. The differential equation (in terms of matrix-vector products) governing the orientation vector, $\vb p$, of axisymmetric spheroidal particles is
\begin{equation}
\dot{\vb p} = \bm\Omega \vb p + \frac{a_1^2 - a_2^2}{a_1^2 + a_2^2}[(\vb I - \vb p \vb p^\T)(\vb E \vb p)],
\end{equation}
where $\bm \Omega = (\vb L - \vb L^\T)/2$ is the antisymmetric vorticity tensor, $\vb E = (\vb L + \vb L^\T)/2$ is the symmetric strain-rate tensor, and $a_1$ and $a_2$ are the radii of the semi-axes that are parallel and orthogonal, respectively, to the axis of axisymmetry. In the limit of infinitely thin sheets, the rate-of-strain prefactor approaches $-1$, which is the value we used in comparing to our numerical simulations.

First, we considered sheets initially oriented with $\phi = 0^\circ$, which corresponds to a flat sheet whose normal is oriented in the flow-gradient plane. These sheets all began initially flat, buckled as they flipped in a way reminiscent of Euler buckling, and eventually flattened out again as they approached the stable flat state along the flow-vorticity plane. Note, again, that such a stable state only exists for infinitely flat sheets oriented in the flow-vorticity plane; sheets with finite thickness would eventually tumble and buckle again periodically, albeit with potentially long periods. Figure \ref{fig:buckle_images} shows snapshots of a particular sheet 
buckling during its trajectory. The degree, or mode, of buckling depended on the bending stiffness and will be discussed more below.

\begin{figure}[h]
\centering
\includegraphics{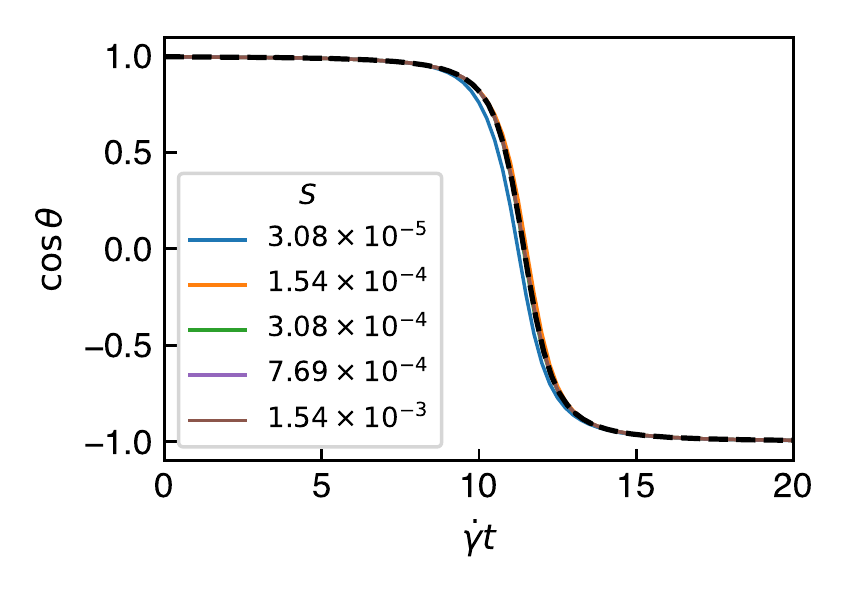}
\caption{Angle formed between the flow-vorticity plane and the line passing through the two ends of a sheet in the flow-gradient plane with initial orientation $(\phi, \theta) = (0^\circ, 5^\circ)$ and for various dimensionless bending rigidities, $S$. The prediction of Jeffery's equations for an infinitely thin oblate spheroid is shown with a black dotted line. See Figure \ref{fig:buckle_images} for snapshots of a (buckling) sheet's orientation evolving in time.}
\label{fig:jeffery}
\end{figure}

\begin{figure*}
\centering
\includegraphics[width=6.5in]{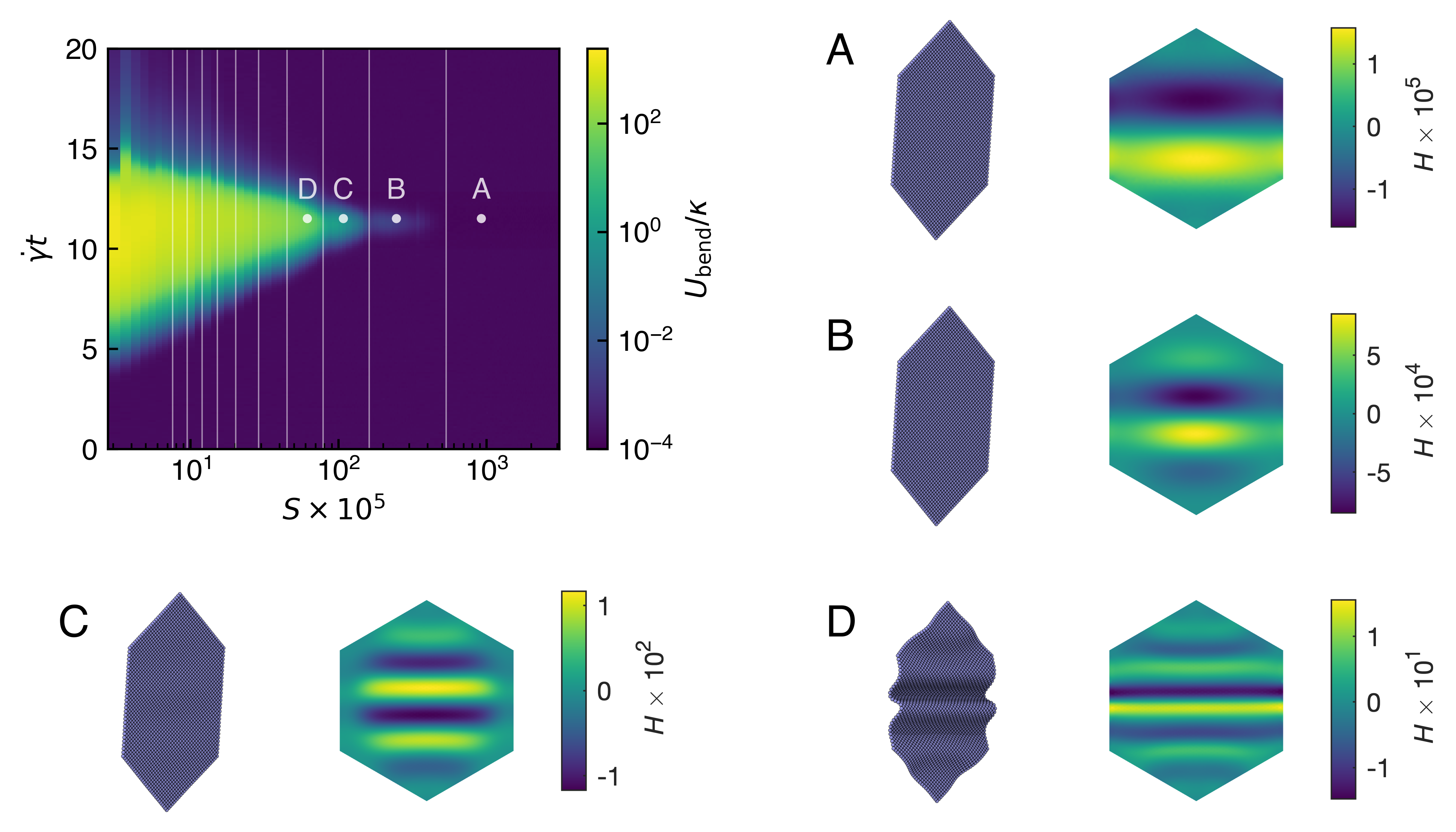}
\caption{Bending energy over time as a function of dimensionless bending rigidity, $S$, for an initial orientation of $(\phi, \theta) = (0^\circ, 5^\circ)$. Vertical lines correspond to the first ten predicted buckling transitions from the quasi-static model at $\theta = \pi/4$, the angle of maximum in-plane stress. Snapshots of sheets with different values of $S$ and at dimensionless time ($\dot \gamma t = 11.5$) corresponding to those of points labeled A-D in the bending energy diagram are shown with (signed) mean curvature, $H$, over the intrinsic coordinates of the sheet.}
\label{fig:buckle_plus_energy}
\end{figure*}

Figure \ref{fig:jeffery} shows the angle between the flow-vorticity plane and the line passing through the two ends of the sheets in the flow-gradient plane as a function of time and $S$, the dimensionless bending rigidity, compared to the angle predicted by the Jeffery orbit of an infinitely thin, rigid, oblate spheroid. This dimensionless bending rigidity --- essentially the sheet's equivalent of the ``elasto-viscous'' number as it is sometimes called in the flexible filament literature \cite{liu2018a} --- is defined as
\begin{equation}
 S = \frac{\kappa }{\pi \eta \dot \gamma L^3}.
 \label{eq:S}
 \end{equation}
Here, $L$ is the characteristic radius of the sheet (equal to 58), and the numerical factor of $1/\pi$ is related to the particular numerical discretization of the sheet (see Appendix). It is clear that the curves are all quite close to the Jeffery trajectory and become closer as bending stiffness increases (i.e., the sheet acts more like a rigid object). That there were not any significant deviations for the smallest value of $S$ is noteworthy considering the fact that that particular sheet buckled to such a large extent that the hard-sphere interactions were engaged. Of course, this is also a symptom of the discretization employed as a finer discretization at the same value of $S$ (i.e., larger number of beads) would allow for higher modes of buckling without hard-sphere contact. Nonetheless, Figure \ref{fig:jeffery} indicates that even sheets that deviate strongly from planarity exhibit a trajectory that can be approximated well by Jeffery's equations, at least for these initial conditions.

\subsection*{Quasi-static elasticity model}

A relatively simple 1D continuum model of elasticity can be used to predict the kind of buckling observed for sheets with initial orientations of $\phi = 0^\circ$. It is quasi-static in the sense that we assume the forces acting upon the sheet are solely a function of the angle $\theta(t)$ and that this angle is determined independently by the Jeffery orbit. This assumption seems to be particularly reasonable in light of the results presented in Figure \ref{fig:jeffery}. Equating moments (see Appendix) and approximating the hydrodynamic stress on the sheet as that of an unperturbed linear flow (i.e., a local analysis), one arrives at the following quasi-static governing equation for small deviations, $w$, of the sheet from planarity along its length, $x$, in the flow-gradient plane:
\begin{equation}
S \dv[2]{\hat x}(f(\hat x) \dv[2]{\hat w}{\hat x}) = q_\perp(\hat x) + q_\parallel(\hat x) \dv{\hat w}{\hat x} + r_\parallel(\hat x) \dv[2]{\hat w}{\hat x}
\label{eq:elasticity}
\end{equation}
where $S$ is the dimensionless bending rigidity from equation \ref{eq:S}, $\hat w = w/L$, $\hat x = (L-x)/L$, and $L$ is the length of the sheet from the center ($\hat x=0$) to the tip ($\hat x=1$). $q_\perp$, $q_\parallel$, and $r_\parallel = -\int_{\hat x}^1 \dd{y} q_\parallel(y)$ are derived from stresses that are perpendicular or parallel to the sheet and, for a hexagonal sheet, are given by:
\begin{align}
f(\hat x) &= \begin{cases}
1 & \hat x < \frac{1}{2} \\
2-2\hat x & \hat x \geq \frac{1}{2}
\end{cases} \\
q_\perp(\hat x) &= f(\hat x)\hat x \sin^2 \theta \qty(\norm{\dot{\vb p}(\theta)}/\dot\gamma - 1) \\
q_\parallel(\hat x) &=f(\hat x) \hat x \sin \theta \cos \theta \\
r_\parallel(\hat x) &= -\sin \theta \cos \theta \begin{cases}
\frac{1}{2}\qty(1 - \hat x^2) - \frac{5}{24}& \hat x < \frac{1}{2} \\
\frac{1}{3} - \hat x^2 + \frac{2}{3}\hat x^3 & \hat x \geq \frac{1}{2}
\end{cases}.
\end{align}
The piecewise nature of these expressions follows from the change in width of the hexagonal sheet along its length in the flow-gradient plane (see Appendix). We imposed the boundary conditions
\begin{equation}
\hat w(0) =\hat  w''(0) = \hat w''(1) = \hat w'''(1) = 0,
\end{equation}
where the primes indicate derivatives. These conditions represent symmetry of the sheet across $\hat x = 0$ and a free end with no applied moment or force, respectively. Similar quasi-static force balances have been applied to model the behavior of filaments subject to large deformations in flow \cite{lindner2015, becker2001, young2007}. Our model differs from these in that it accounts for heterogeneity of the hydrodynamic and elastic forces along the sheet in the flow-gradient plane and neglects the Lagrange multipliers typically employed to enforce inextensibility.
Because the model is meant to identify the first deviations from flatness of the sheet only, as done in the classical analysis of Euler buckling, we believe these approximations are appropriate.
It is also worth mentioning that, while this work considers hexagonal sheets, the above equations could be altered in a straightforward manner to model sheets of other shapes via modification of the function $f$.

\begin{table}[h]
\centering
\small\caption{Predicted values of $S$ and mode shapes corresponding to buckling modes of a hexagonal sheet with orientation $\phi = 0^\circ$ at the angle of maximum in-plane stress, $\theta = \pi/4$.}
\label{tab:S_values}

\begin{tabular}{@{}rccccc@{}}
\toprule
$S \times 10^5$ & 534.92 & 161.20 & 78.93 & 45.09 & 28.88 \\ 
Mode & \includegraphics[scale=0.4]{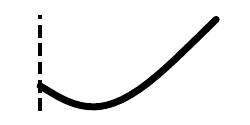} & \includegraphics[scale=0.4]{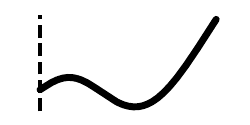} & \includegraphics[scale=0.4]{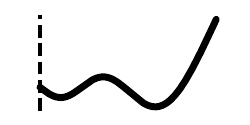} & \includegraphics[scale=0.4]{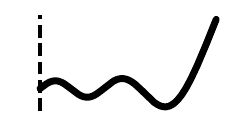} & \includegraphics[scale=0.4]{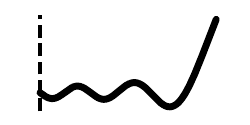} \\
\bottomrule 
\end{tabular}

\end{table}

Unlike Euler buckling, equation \ref{eq:elasticity} features an inhomogeneous term due to stresses perpendicular to the sheet. However, the solution to equation \ref{eq:elasticity} can easily be split into a homogeneous and particular solution as $\hat w = \hat w^h + \hat w^p$, indicating that it is the \textit{homogeneous} part that is responsible for buckling instabilities. The model can be solved via a basis function expansion in ultraspherical polynomials \cite{olver2013} as
\begin{equation}
\vb A \vb w^h = S^{-1} \vb B \vb w^h,
\label{eq:eigenvalue}
\end{equation}
where $\vb A$ represents the fourth-order derivative operator, $\vb B$ represents the right-hand side \footnote{The piecewise nature of equation \ref{eq:elasticity} was not treated specially. Smooth polynomials of high degree can approximate piecewise continuous functions with minimal error \cite{trefethen2013}.} of equation \ref{eq:elasticity}, and $\vb w^h$ is a vector containing the coefficients of the polynomials. The boundary conditions are contained in the top four rows of $\vb A$, and the top four rows of $\vb B$ are accordingly set to 0. Equation \ref{eq:eigenvalue} represents a generalized eigenvalue problem, where $S$ (or $S^{-1}$) is a generalized eigenvalue. Solving equation \ref{eq:eigenvalue} with a basis of 100 polynomials yielded a series of buckling modes and buckling eigenvalues. The first five of these eigenvalue-mode pairs can be seen in Table \ref{tab:S_values}.

\begin{figure}[ht]
\centering
\includegraphics{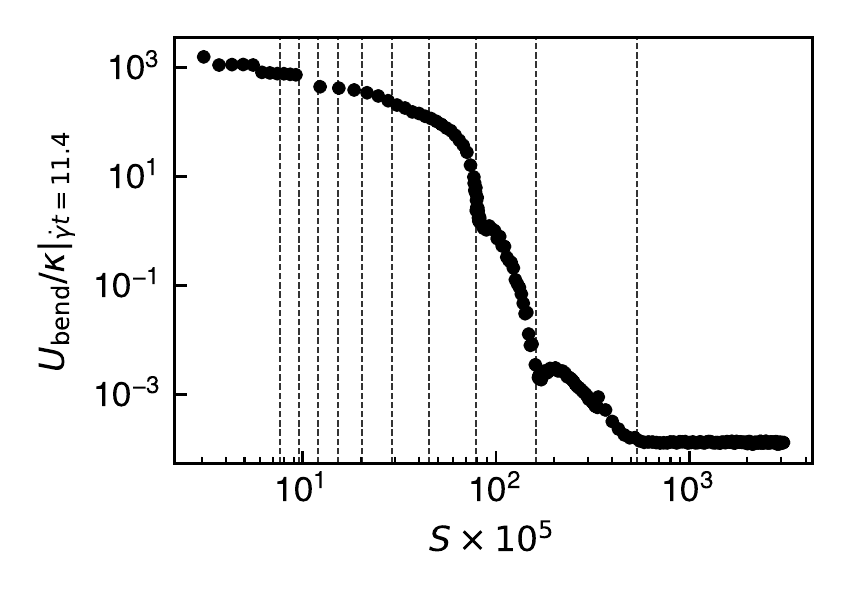}
\caption{Bending energy as a function of dimensionless bending rigidity, $S$, for an initial orientation of $(\phi, \theta) = (0^\circ, 5^\circ)$ at $\dot \gamma t = 11.4$, the point at which sheets are oriented vertically to the flow. Vertical lines correspond to the first ten predicted buckling transitions from the quasi-static model at $\theta = \pi/4$, the angle of maximum in-plane stress.}
\label{fig:bend_max_energy}
\end{figure}

\begin{figure*}[ht]
\centering
\includegraphics{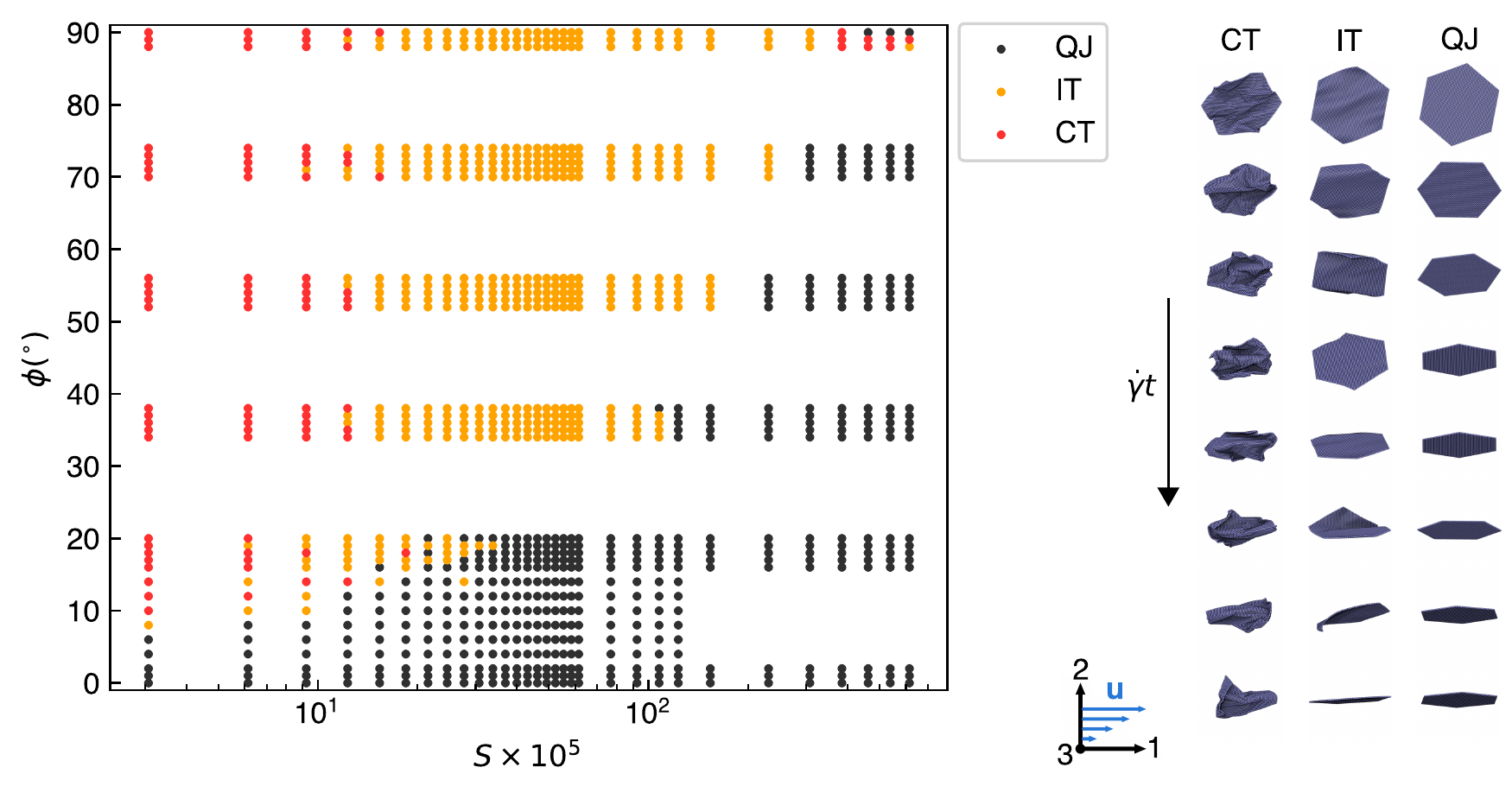}
\caption{\textbf{Left)} Dynamical states of the sheet as a function of dimensionless bending rigidity, $S$, and initial angle $\phi$. ``QJ'' denotes ``quasi-Jeffery'' (close to a Jeffery orbit), ``IT'' denotes ``initial tumbling'' (eventually reaching the stable flat state by $\dot \gamma t = 1000$), and ``CT'' denotes ``continuous tumbling''. \textbf{Right)} Example snapshots (every $\dot \gamma t = 12.5$ units of time) of continuously tumbling, initially tumbling, and quasi-Jeffery trajectories with initial orientation $\phi = 54^\circ$ and $S$ values of $3.08 \times 10^{-5}$, $6.15 \times 10^{-4}$, and $3.08 \times 10^{-3}$, respectively. The flow (1), gradient (2), and vorticity (3) directions are depicted at the bottom. Movies of representative trajectories are included in the SI.}
\label{fig:chaos_phase}
\end{figure*}

The success of this simple 1D model for sheets with initial orientations of $\phi = 0^\circ$ and $\theta = 5^\circ$ can be seen Figure \ref{fig:buckle_plus_energy}. The maximum in-plane stress is exerted on a flat sheet at an angle of $\theta = \pi/4$. Neglecting transient effects, we used this value of $\theta$ to predict the growth of buckling modes before they ultimately decayed as the sheet reached the stable flat state at long times. Indeed, these values of $S$ seem to faithfully predict the different buckling regimes observed. Sheets with varying dimensionless bending rigidity exhibited different modes of buckling at the apex of their trajectories (i.e., when they were most vertical and the point in time at which bending energy was approximately maximized). Some snapshots of different sheets within these different buckling regimes are shown in Figure \ref{fig:buckle_plus_energy} along with images of their mean curvature, $H$, which was calculated with the ``cotangent'' formula that is commonly used in computational geometry \cite{meyer2003}. The different modes attained are clearly apparent in these mean curvature diagrams.

The bending energy, or the maximal bending energy, as a function of $S$ shown in Figure \ref{fig:bend_max_energy} is quite nonlinear and is inherently not described by this 1D quasi-static model. In fact, for the same reason, it is somewhat remarkable that the model \textit{does} capture the different buckling modes of such a complex dynamical system. Some of this complicated dependence on $S$ for the smallest values explored ($S < 5\times 10^{-5}$), though, can be attributed to the interplay of buckling, strong hydrodynamic interactions, and self-avoidance via hard-sphere interactions, which, along with the smallest length scale of the discretization, places a bound on attainable bending energies. However, for values of $S$ above the first predicted buckling transition ($S = 5.35 \times 10^{-3}$), the bending energy scales linearly with $\kappa$, as sheets do not buckle and are only perturbed slightly away from the flat state.

It is also interesting to note that the process of flattening out after $\dot \gamma t \approx 11.4$ proceeded via 2D modes of motion which likely cannot be well described by any 1D model (see Movie S2, for example). This lack of time reversal symmetry is distinct from Jeffery orbits of rigid spheroidal particles, which are invariant under the transformation $(t, \vb L) \mapsto (-t, -\vb L)$, and is also exhibited by the bending energy diagram in Figure \ref{fig:buckle_plus_energy}, for bending energy \textit{decreases} more quickly during flattening than it \textit{increases} during buckling. 

\section{Chaos and Tumbling}

\begin{figure*}[ht]
\centering
\includegraphics{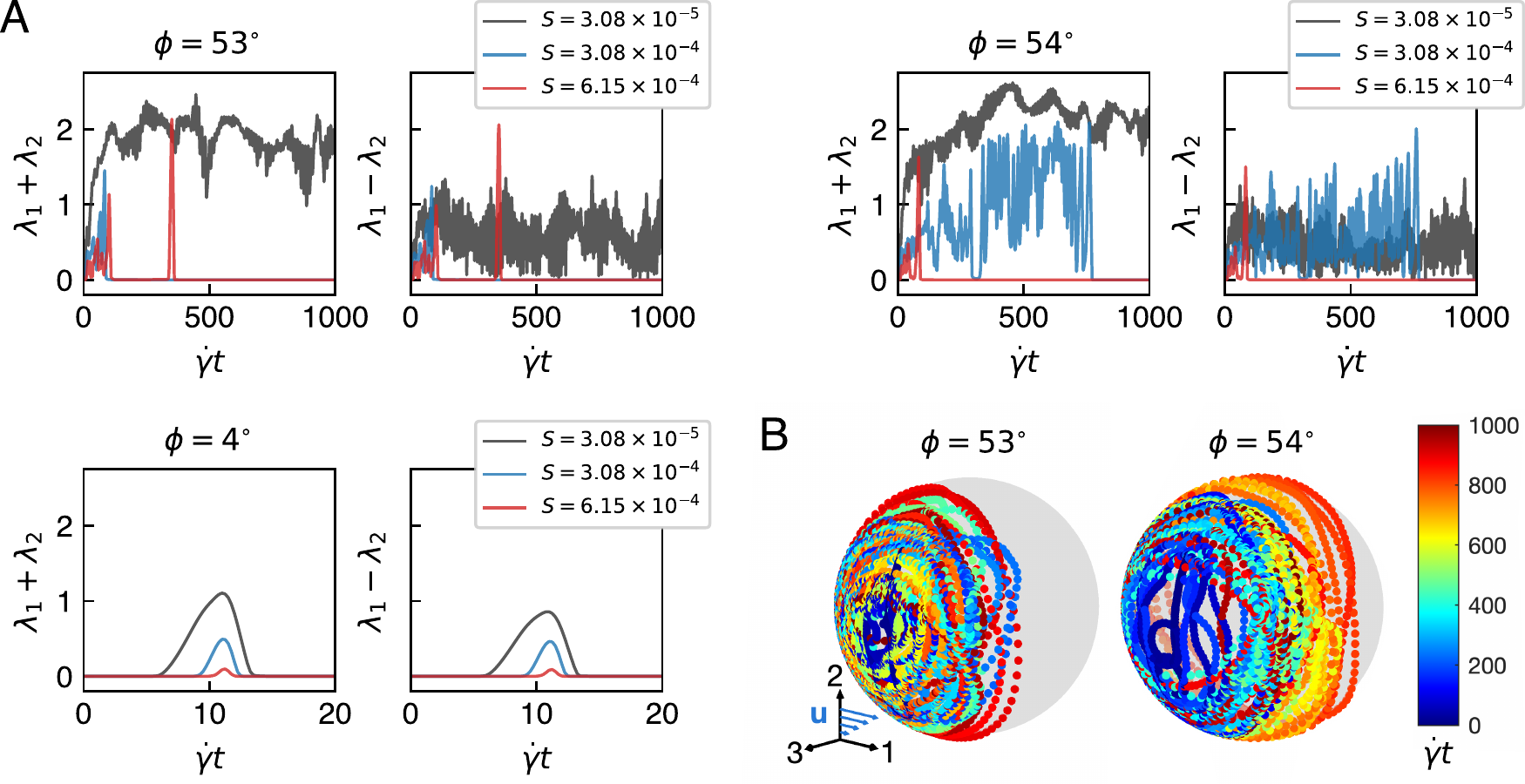}
\caption{Crumpling and chaos in some representative examples of sheets. \textbf{A)} The sum and difference of the eigenvalues, $\lambda_1$ and $\lambda_2$, of the orientational covariance matrix, $\bar{\vb C}$, for three different values of dimensionless bending rigidity, $S$, and three different initial orientation angles, $\phi$. \textbf{B)} Mean orientation as a function of time (represented by color) for $S = 3.08 \times 10^{-5}$ at slightly different initial orientation angles. The flow (1), gradient (2), and vorticity (3) directions are depicted at the bottom. A small change in initial orientation results in drastically different aperiodic trajectories, which is the hallmark of chaos.}
\label{fig:mean_covariance}
\end{figure*}

In experimental systems, sheets may not be perfectly oriented with $\phi=0^\circ$, so we sought to explore the dynamical behavior of sheets with different initial orientations $\phi$ ranging from $0^\circ$ to $90^\circ$. To quantify such dynamical behavior, we focused on two summary statistics: the \textit{mean orientation} and the \textit{orientational covariance matrix}. The normal vector at each point of a regular surface lives on the unit sphere, $S^2$. Thus, the mean orientation of a sheet can be calculated via the weighted Fréchet mean \cite{pennec2006},
\begin{equation}
\bar{\vb n} = \argmin_{\vb x \in S^2} \sum_{i \in \{\bigtriangleup\}} w_i \dist(\vb x, \vu n_i)^2,
\end{equation}
where the sum is over all triangles (i.e., all groups of three close-packed beads) of the mesh, $\vu n_i$ is the normal of triangle $i$, the weight $w_i$ is the area of triangle $i$, and $\dist : (\vb x, \vb y) \mapsto \arccos(\vb x \vdot \vb y)$ is the Riemannian distance between two points on the sphere. Note that it is important to ensure that the normals are consistently oriented across the whole mesh. Additionally, with these choice of weights, the sum serves as a kind of finite-volume approximation of the continuous integral over a smooth, continuous surface. Like the typical arithmetic mean in Euclidean space, the Fréchet mean is a point on a manifold (in this case the unit sphere) that minimizes the squared distance to a set of points. The (weighted) orientational covariance matrix is given by:
\begin{equation}
\bar{\vb C} = \frac{\sum_j w_j}{\qty( \sum_j w_j )^2 - \sum_j w_j^2} \sum_{i \in \{\bigtriangleup\}} w_i \log_{\bar{\vb n}}(\vu n_i) \log_{\bar{\vb n}}(\vu n_i)^\T
\end{equation}
where $\log_{\vb x}: S^2 \to T_{\vb x} S^2$ is the inverse of the exponential map and maps points on the sphere onto the tangent plane at $\vb x$. Specifically, $\log_{\vb x}$ is calculated as
\begin{equation}
\log_{\vb x}: \vu n \mapsto \dist(\vb x, \vu n) \frac{(\vb I - \vb x \vb x^\T) (\vu n - \vb x)}{\norm{(\vb I - \vb x \vb x^\T) (\vu n - \vb x)}_2},
\end{equation}
where $(\vb I - \vb x \vb x^\T) (\vu n - \vb x)$ represents an orthogonal projection of the vector $(\vu n - \vb x)$ in Euclidean space (\textit{not} intrinsically on the sphere).
One can think of this process visually as unwrapping the path on the sphere between two points like an inextensible string onto the tangent plane of the first point, all while maintaining the string's original direction. The covariance matrix $\bar{\vb C}$ represents the spread of the orientations of the sheet about its mean orientation. A sheet that is flat would exhibit zero spread since all the normals would be identical, whereas a crumpled sheet would exhibit quite a bit of variance about the mean orientation since the normals across the sheet would point in different directions \footnote{There is some ambiguity if the sheet curls around itself more than $180^\circ$ since the Riemannian distance would not reflect this (i.e., distances between two points on the sphere are bounded between $0$ and $\pi$). Such a scenario, though, is only possible with highly crumpled sheets, which would still be reflected in these statistics as a large variance about the mean orientation.}.

\begin{figure}[ht]
\centering
\includegraphics{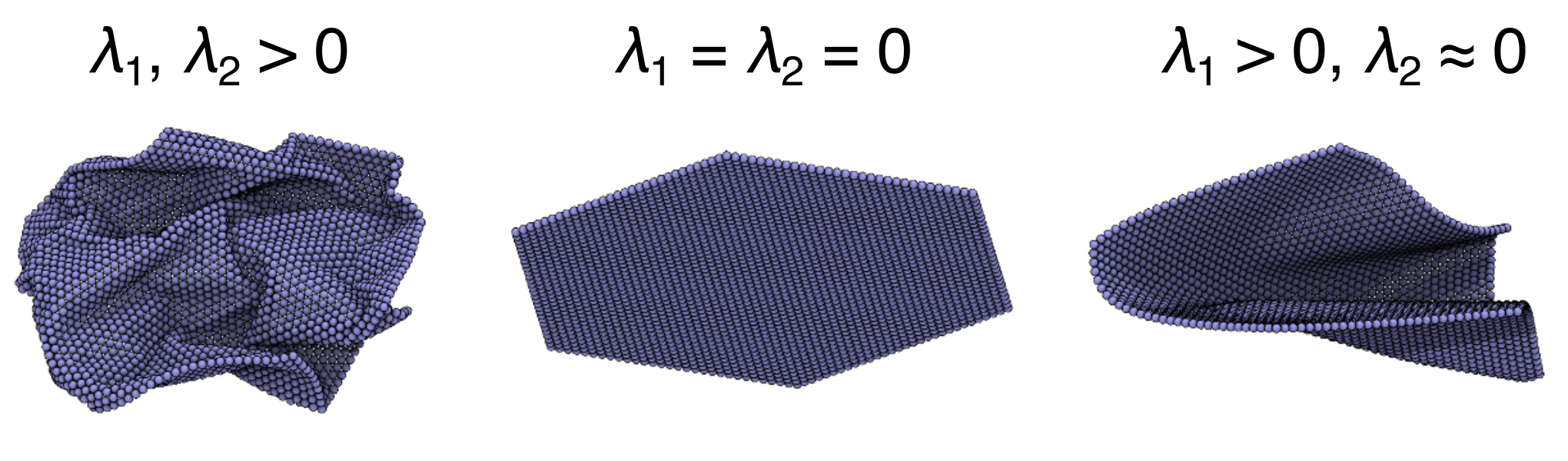}
\caption{Illustration of the physical meaning of the eigenvalues of the orientational covariance matrix $\bar{\vb C}$. An isotropically crumpled sheet is associated with two positive eigenvalues, a perfectly flat sheet is associated with two zero eigenvalues, and an anisotropically creased sheet is associated with one positive eigenvalue and another close to zero.}
\label{fig:lambda_illustration}
\end{figure}

Many numerical simulations of long duration (up to $\dot \gamma t = 1000$) were conducted with sheets of varying bending rigidities and initial angles $\phi$. Several different dynamical behaviors were observed, as shown in Figure \ref{fig:chaos_phase}. Movies of representative trajectories are also included in the SI. Some sheets (shown with black dots) very closely tracked the trajectory predicted by the Jeffery orbit of an infinitely thin spheroid with an equivalent initial orientation. Specifically, we denote sheet trajectories as ``quasi-Jeffery'' if $\abs{\bar{\vb n}(t) \vdot \vb p(t)} > 0.99$ for all times $t$ (i.e., the mean normal is sufficiently close to the Jeffery orbit). Unsurprisingly, stiffer sheets (higher $S$ values) exhibit a wider range of initial orientations that result in quasi-Jeffery trajectories. It is also important to note that the cutoff value of $0.99$ did not significantly affect labeling. In other words, sheets that deviated from Jeffery orbits did so considerably. Although not explored in this work, we believe that the 2D buckling patterns exhibited by quasi-Jeffery sheets with $\phi \neq 0^\circ$ would likely be described well by a similar 2D quasi-static elasticity model.

Sheets that were not quasi-Jeffery but eventually reached the stable flat state by $\dot \gamma t = 1000$ were labeled in Figure \ref{fig:chaos_phase} as ``initial tumbling''. The rest of the sheets, which never reached the stable flat state, were denoted as ``continuous tumbling''. Among the sheets that were labeled as initial tumbling, some exhibited erratic tumbling at the beginning of the trajectory while others flipped after long periods of time of ostensibly approaching the flat state (see $S=6.15\times 10^{-4}$ at $\phi = 53^\circ$ in Figure \ref{fig:mean_covariance} as an example). Therefore, this ``initial tumbling'' regime can be viewed as a transitional regime between quasi-Jeffery and continuously tumbling trajectories. Some apparent outliers can be found in top-right corner of Figure \ref{fig:chaos_phase} for large values of $S$ and values of $\phi$ near $90^\circ$. Some of these sheets oscillated back and forth in the flow seemingly indefinitely and in a stable manner. However, we believe that a duration of $\dot \gamma t = 1000$ was not long enough to observe what would likely be a return to the stable flat state. That quasi-Jeffery orbits occurred for the inherently unstable initial orientation of $\phi = 90^\circ$ around a value of $S$ that corresponds to the first predicted buckling transition at $\phi = 0^\circ$ is likely not coincidental and warrants further investigation.

The sheets that did tumble did so in a chaotic way without any regular periodicity. This behavior is similar to that observed with flexible filaments subjected to shear flow \cite{slowicka2015,slowicka2020,lagrone2019}. Figure \ref{fig:mean_covariance} shows summary statistics for representative sheets of various values of $S$ with similar initial angles $\phi$. Namely, the eigenvalues, $\lambda_1$ and $\lambda_2$, of the orientational covariance matrix are presented since they are invariant with respect to rotations or a change of basis. Their sum (equal to $\tr(\bar{\vb C})$) is representative of the total variance of normals across the surface of the sheet, and their difference is representative of the degree of anisotropy of the distribution. Figure \ref{fig:lambda_illustration} illustrates the physical meaning of these eigenvalues as they relate to the geometry of the sheet. For example, moderately uniform crumpling can be seen for the softest sheet (black curves) at $\phi = 53^\circ$ and $\phi = 54^\circ$ in Figure \ref{fig:mean_covariance}. Additionally, the isolated spike during the initially tumbling trajectory of the stiffest sheet (red curve) at $\phi = 53^\circ$ in Figure \ref{fig:mean_covariance} represents a highly anisotropic transient crease that occurs as the sheet flips over itself before settling into the flat state.

Several conclusions can be gleaned from the data presented in Figure \ref{fig:mean_covariance}. First, it is clear that softer sheets (smaller $S$ values) generally exhibit greater orientational variance than stiffer sheets, both for tumbling trajectories and for the quasi-Jeffery trajectory ($\phi = 4^\circ$) presented. This trend is entirely expected: softer sheets manifest greater degrees of crumpling, which directly corresponds to a greater variance of normals about the mean orientation. Second, it is apparent that the difference of the eigenvalues, $\lambda_1 - \lambda_2$, as a fraction of their sum was in general \textit{smallest} for the \textit{softest} sheet. This behavior indicates that the soft sheet tends to crumple more isotropically (similar crumpling in all directions), whereas stiffer sheets crumple more anisotropically (like folding a crease along a single direction). Finally, it is clear for $\phi = 53^\circ$ and $\phi = 54^\circ$ --- the angles that produced tumbling --- that a small change in initial conditions yielded drastically different trajectories, as evidenced by the pattern of the orientational covariance and the mean orientations themselves (panel B of Figure \ref{fig:mean_covariance}). In fact, these continuously tumbling trajectories diverged away from each other exponentially quickly. This, of course, is the classical signature of chaos (see Figure \ref{fig:lyapunov} and the Appendix for Lyapunov exponent estimates). It is important to note that although the sheets tumble in a circular way and the covariance seems to ``ebb and flow'' over time, we did not detect any significant signature of regular periodicity in power spectral densities. One can also see that for the sheets that \textit{initially} tumble, the time at which the stable flat state is attained is quite erratic. Overall, the trends discussed for these few sheets hold for the many others that were studied and depicted as single dots in the state diagram of Figure \ref{fig:chaos_phase}.

Interestingly, chaotic trajectories are often associated with crumpled configurations (i.e., higher orientational variances). This connection is not obvious \textit{a priori}. Perhaps it is the more crumpled states of softer sheets that allow stronger and more complex fluid-bead and bead-bead interactions to occur and give rise to sensitive, chaotic dynamics.

\section{Conclusions}

The dynamical behavior of athermal 2D sheets immersed in a shear flow at low Reynolds number was investigated via immersed boundary simulations with a model semiflexible sheet at high Föppl-von Kármán numbers (i.e., softer bending relative to stretching modes of motion). The main governing dimensionless parameter of the system is the dimensionless bending rigidity, $S = \kappa /(\pi \eta \dot \gamma L^3)$. Our findings on the behavior of athermal sheets can be summarized succinctly as follows:
\begin{enumerate}
\item For flat sheets initially oriented with a normal in the flow-gradient plane, transient buckling occurs and can be predicted as a function of $S$ quite accurately using a simple 1D elasticity model.
\item Smaller values of $S$ can result in chaotic, continuously tumbling trajectories, but not for all initial orientations.
\item Chaotic trajectories are associated with crumpled conformations.
\item Sheets that do not tumble chaotically (generally those lying near the flow-vorticity plane or large values of $S$) are described well by the equivalent Jeffery orbit of a rigid, spheroidal particle.
\end{enumerate}
This delineation of the dynamical regimes of athermal sheets should inform the design of solution processing protocols of flexible 2D materials, where crumpling or buckling may or may not be desired for different applications. Specifically, future directions and applications include better understanding the influence of shear-induced morphological changes of dispersed nanosheets on bulk rheological properties (e.g., shear-dependent viscosity), the potential migration of the sheets in unbounded shear or in other flow geometries, and the design of precision flow systems to tune nanomaterial conformations. Additionally, the use of the ``intrinsic'' summary statistics employed in this work (mean orientation and the orientational covariance matrix) may prove to be useful in future theoretical and experimental studies of related systems.

\begin{acknowledgments}
K.S.S. was supported by the Department of Energy Computational Science Graduate Fellowship program under grant No. DE-FG02-97ER25308. M.S.S. was supported by the Department of Energy, Office of Science, Basic Energy Sciences under grant No. DE-FG02-08ER46488 Mod 0008. J.W.S. was supported by NSF Career Award No. CBET-1554398. The authors would like to thank M. Ekiel-Jeżewska, G. McKinley, P. Doyle, and N. Fakhri for helpful discussions.
\end{acknowledgments}

\bibliography{extracted.bib}

\clearpage

\onecolumngrid
\appendix
\section{Derivation of Quasi-Static Elasticity Model}

\begin{figure}[ht]
\centering
\includegraphics[width=3.25in]{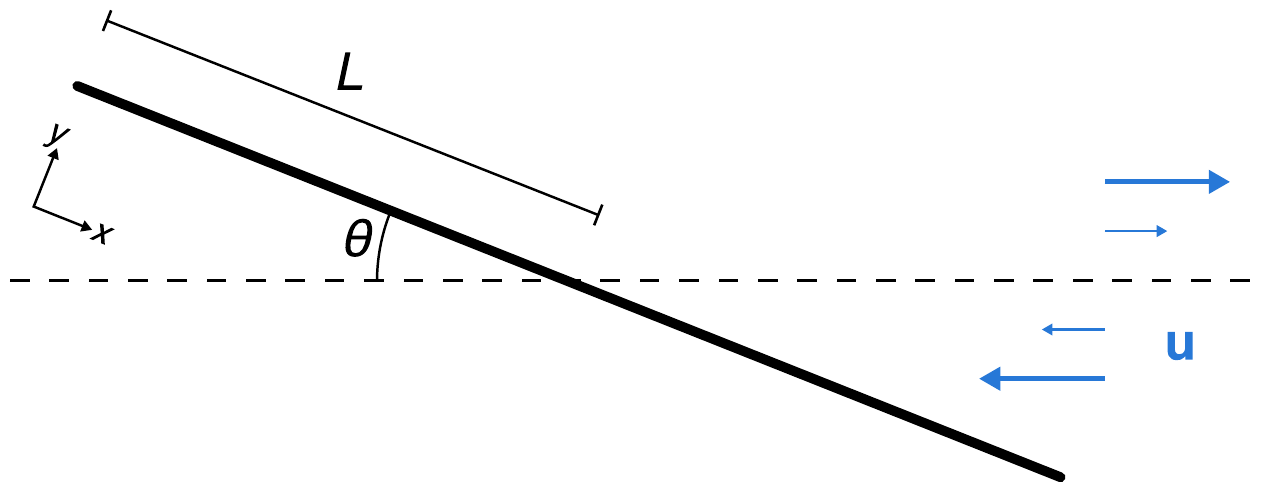}
\caption{Schematic of a rod (or sheet) in shear flow.}
\label{fig:elasticity_diagram}
\end{figure}

First, let us consider a 1D rod (see Figure \ref{fig:elasticity_diagram}) under the assumptions of classical Euler-Bernoulli beam theory with a flexural rigidity of $EI$, where $E$ is the 3D Young's modulus and $I$ is the second moment of area of the cross-section \cite{landau2008}. The equation for a small deflection, $w$, of the rod in the $y$ direction is given by
\begin{equation}
EI \dv[2]{w}{x} = -\int_0^x \dd{x'} q_y(x')(x - x') - \int_0^x \dd{x'} q_x(x')\qty[ w(x) - w(x') ],
\end{equation}
where the right-hand side is given by the moments on the rod in the perpendicular and parallel directions (similar to a self-buckling analysis) and the functions $q$ are given by
\begin{align}
q_x &= \alpha \eta \dot \gamma \cos(\theta) (L-x) \sin(\theta) \\
q_y &= \alpha \eta \qty(\dot \gamma - \norm{\dot{\vb p}} ) \sin(\theta) (L-x) \sin(\theta).
\end{align}
Here, it is assumed that the force per unit length is proportional to the viscosity, $\eta$, and the velocity of an unperturbed linear shear flow with a dimensionless constant $\alpha$ that depends on the exact shape of the rod. The angle $\theta$ is approximated independently by the equivalent Jeffery orbit, $\vb p$, of a rigid object, which is also used to adjust the net moment in the perpendicular direction. In this analysis, it is also assumed that there is no transient response, hence the ``quasi-static'' nature of the model.

Differentiation once with respect to $x$ yields
\begin{equation}
\dv{x}(EI \dv[2]{w}{x}) = -\int_0^x \dd{x'} q_y(x') - \dv{w}{x} \int_0^x \dd{x'} q_x(x'),
\end{equation}
and differentiating once more yields
\begin{equation}
\dv[2]{x}(EI \dv[2]{w}{x}) = -q_y(x) - q_x(x) \dv{w}{x} - \dv[2]{w}{x} \int_0^x \dd{x'} q_x(x').
\end{equation}
The boundary conditions are
\begin{align}
\left. \dv[3]{w}{x} \right|_0 = \left. \dv[2]{w}{x} \right|_0 &= 0 \\
\left. \dv[2]{w}{x} \right|_L = \left. w \right|_L &= 0.
\end{align}
The first two represent the free end of the rod at $x=0$, which is subjected to no moment or force. The latter two reflect the symmetry of the behavior of the rod about the center streamline.

Nondimensionalizing the equation using $\hat w = w/L$ and $\hat x = (L-x)/L$ yields:
\begin{equation}
B \dv[4]{\hat w}{\hat x} = \sin(\theta) \left[ \hat x \sin\theta \qty( \frac{\norm{\dot{\vb p}}}{\dot \gamma} - 1) + \hat x \cos(\theta) \dv{\hat w}{\hat x}  - \frac{\cos(\theta)}{2}\qty(1 - \hat x^2) \dv[2]{\hat w}{\hat x} \right],
\end{equation}
where
\begin{equation}
B = \frac{EI}{\alpha \eta \dot \gamma L^4}
\end{equation}
is the elasto-viscous number \cite{liu2018a}.

For a 2D sheet extending in and out of the plane of the paper (the $z$ direction), deforming purely in a 1D way, and experiencing stresses independent of the $z$ direction, $EI$ can be considered an effective modulus proportional to $\int \dd{z} \kappa$, where $\kappa$ is the bending rigidity of the 2D sheet. Likewise, $q_x$ and $q_y$ can be considered as effective 1D stresses where $\alpha \propto \int \dd{z} \beta$ for some constant $\beta$ (with units of length$^{-1}$) that depends on the hydrodynamics of the sheet. Thus, the relevant dimensionless quantity becomes $S$, defined as
\begin{equation}
S = \frac{\kappa}{\alpha \eta \dot \gamma L^3},
\end{equation}
instead of $B$. It is interesting to note that for varying widths, $W$, in the $z$ direction, $S$ scales like $W/L^4$. For widths approaching the bead radius, the elasto-viscous $L^4$ scaling for filaments is recovered.

\subsection{Determination of Constants for Bead Model}
First, as mentioned in the main text, the bending rigidity for a triangulated sheet with dihedral forces is $\sqrt{3}$ times greater than the bending rigidity of the equivalent continuum sheet \cite{gompper1996}. Thus, in mapping $S$ to simulation data, the numerator of $S$ should be multiplied by $\sqrt{3}$. Additionally, one can calculate $\beta$ based on the discretization of beads employed. Since the beads (of radius $a=1$) are 2D close-packed, the stress on the sheet due to a velocity, $u$, in a given area, $A$, can be approximated as
\begin{equation}
\text{Stress} = \frac{\text{Force}}{A} = \frac{6\pi\eta a u N_\text{beads}}{A} = \frac{6\pi\eta a u \frac{\sqrt{3}\pi}{6}\frac{A}{\pi a^2}}{A} = \frac{6\pi \eta \sqrt{3}}{6 a} u,
\end{equation}
where $\sqrt{3} \pi / 6$ is the density of close-packed spheres in 2D. Therefore, $\beta$ for our model must be $6\pi \sqrt{3} / (6a)$, which does indeed have units of length$^{-1}$. The integral of $\beta$ contributes a factor of $a/L$ to the definition of $S$. However, $\int \dd{z} \kappa \sim L(L/a) \kappa$, where the proportionality factor of $L/a$ reflects the density of dihedrals on the discrete sheet and derives from the fact that there are $(2L/2a)$ bonds in the $z$ direction. This factor of $L/a$ cancels out the contribution of $a/L$ from $\beta$. With these considerations, $S$ is independent of the bead length scale $a$ and is given by
\begin{equation}
S = \frac{6\kappa}{6\pi \eta \dot \gamma L^3}.
\end{equation}

\subsection{Accounting for Changing Width} As one travels along the $x$ direction, the width of a hexagonal sheet in the $z$ direction changes. That is, the width of a hexagonal sheet as a fraction of its total width is given by
\begin{equation}
f(x) = \begin{cases}
\frac{2x}{L} & x < \frac{L}{2} \\
1 & x \geq \frac{L}{2}
\end{cases}.
\end{equation}
One can account for this change of width in the effective 1D stress constant $\alpha$ and flexural rigidity $EI$ by simply multiplying $\beta$ and $\kappa$ respectively by this (dimensionless) piecewise function, rendering both quantities $x$-dependent. The stresses $q_y$ and $q_x$, in turn, gain additional $x$ dependence due to this changing width. Now,
\begin{align}
q_x &= \alpha f(x) \eta \dot \gamma \cos(\theta) (L-x) \sin(\theta) \\
q_y &= \alpha f(x) \eta \qty(\dot \gamma - \norm{\dot{\vb p}} ) \sin(\theta) (L-x) \sin(\theta).
\end{align}
It is these expressions that lead to equations 9-12 in the main text.

\begin{figure}[h]
\centering
\includegraphics{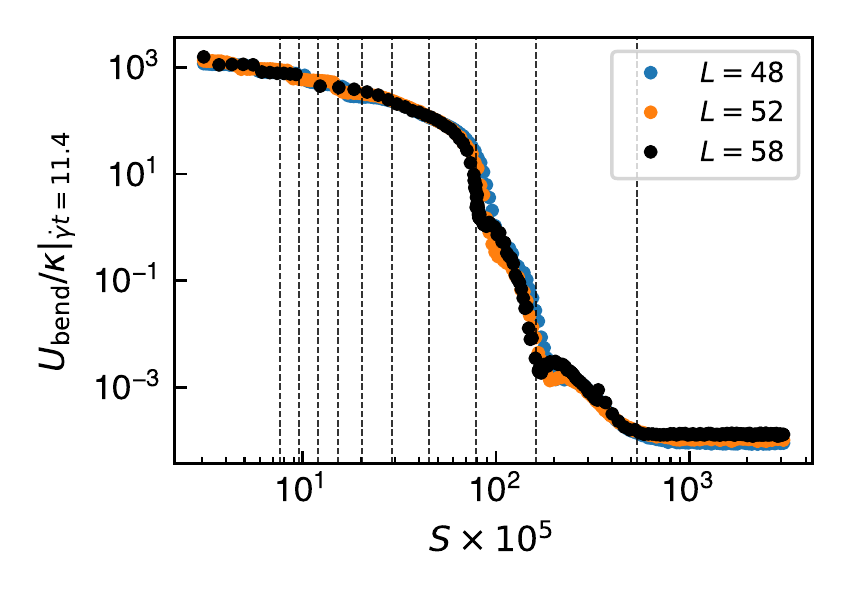}
\caption{Bending energy as a function of dimensionless bending rigidity, $S$, for an initial orientation of $\phi = 0^\circ$ and $\theta = 5^\circ$ at $\dot \gamma t = 11.4$, the point at which sheets are oriented vertically to the flow. The different colors represent differently sized sheets in the simulation. Vertical lines correspond to the first ten predicted buckling transitions from the quasi-static model at $\theta = \pi/4$, the angle of maximum in-plane stress.}
\label{fig:bend_max_energy_plus}
\end{figure}

\subsection{Verification of Scaling}

Figure \ref{fig:bend_max_energy_plus} is essentially the same as Figure \ref{fig:bend_max_energy} but shows the bending energy attained when vertically oriented for differently sized sheets of lengths $L = 48$, $52$, and $58$ (with numbers of beads equal to 1801, 2107, and 2611, respectively). In these additional simulations, the bond stiffness was varied in order to maintain a constant FvK number. As one can see, the proposed $L^3$ scaling of the dimensionless bending rigidity, $S$, indeed holds as all of the curves collapse onto each other.

\begin{figure*}[ht]
\centering
\includegraphics{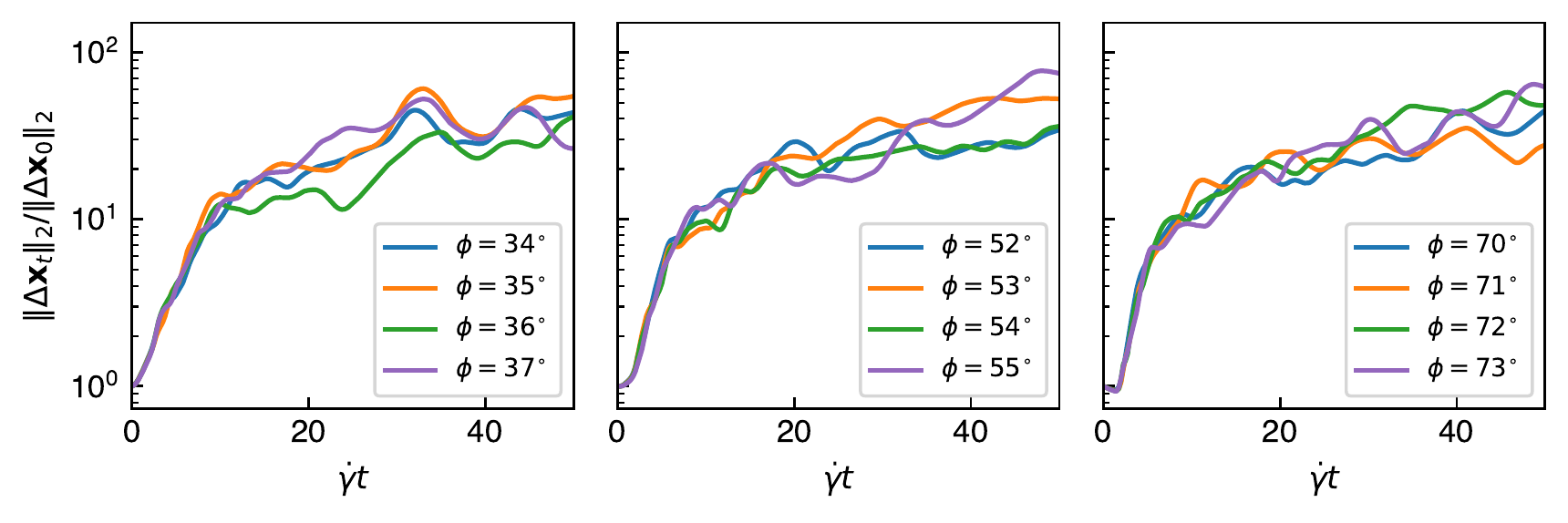}
\caption{The norm of the difference between the vertex coordinates of two continuously tumbling sheets with initial conditions of $(\phi, \theta = 5^\circ)$ and $(\phi + 1^\circ, \theta = 5^\circ)$ over time. For all trajectories shown, $S = 3.08 \times 10^{-5}$. An exponential divergence in trajectories can be seen for short times.}
\label{fig:lyapunov}
\end{figure*}

\section{Maximal Lyapunov Exponents}

Figure \ref{fig:lyapunov} shows the norm of the difference between the vertex coordinates of sheets of dimensionless rigidity $S = 3.08 \times 10^{-5}$ with slightly different initial conditions (i.e., where $\phi$ differs by $1^\circ$). This value of $S$ corresponds to trajectories that are continuously tumbling for all of the orientations featured in Figure \ref{fig:lyapunov}. At each timestep, the coordinates were translated in space so that the center of mass of the sheet was located at the origin. Given that sheets do not stretch to a significant extent, this translation ensures that the coordinates of the sheet live in a compact state space. One can see that, at least for short times, the small perturbation in initial orientation $\phi$ causes trajectories to diverge from each other exponentially. Maximal Lyapunov exponents \cite{strogatz2001} can be estimated by fitting the data to the functional form
\begin{equation}
\norm{\Delta \vb x_t}_2 = \norm{\Delta \vb x_0}_2 e^{\lambda_\mathrm{max} \dot \gamma t}.
\end{equation}
Here, $\Delta \vb x_t$ is the difference between the vertex coordinates (after subtracting the center of mass) of two sheets with different initial conditions at time $t$. Performing a least-squares linear regression on the logarithm of the norm differences at times $2.25 \leq \dot \gamma t \leq 4.75$ for the three groups of initial angles shown in Figure \ref{fig:lyapunov} yields maximal Lyapunov exponents of 0.31, 0.41, and 0.56 for $\phi$ near $36^\circ$, $54^\circ$, and $72^\circ$, respectively. Thus, the maximal Lyapunov exponents are dependent on initial orientation and increase as $\phi$ increases. It should be noted that these Lyapunov exponents are dimensionless; in real units of time, one can see that they scale with the shear rate.


\section{Movies}

Movies can be found at \href{https://dx.doi.org/10.6084/m9.figshare.13238447}{doi:10.6084/m9.figshare.13238447} and include the following:

\begin{enumerate}
\item{Buckling of a sheet with $S = 3.08 \times 10^{-5}$ and initial orientation $\phi = 0^\circ$.}
\item{Quasi-Jeffery trajectory of a sheet with $S = 3.08 \times 10^{-5}$ and initial orientation $\phi = 4^\circ$.}
\item{Continuously tumbling trajectory of a sheet with $S = 3.08 \times 10^{-5}$ and initial orientation $\phi = 54^\circ$.}
\item{Buckling of a sheet with $S = 6.15 \times 10^{-4}$ and initial orientation $\phi = 0^\circ$.}
\item{Quasi-Jeffery trajectory of a sheet with $S = 6.15 \times 10^{-4}$ and initial orientation $\phi = 4^\circ$.}
\item{Initially tumbling trajectory of a sheet with $S = 6.15 \times 10^{-4}$ and initial orientation $\phi = 54^\circ$.}
\item{Quasi-Jeffery trajectory (and imperceptible buckling) of a sheet with $S = 3.08 \times 10^{-3}$ and initial orientation $\phi = 0^\circ$.}
\item{Quasi-Jeffery trajectory of a sheet with $S = 3.08 \times 10^{-3}$ and initial orientation $\phi = 54^\circ$.}
\end{enumerate}

\end{document}